\documentclass[pdflatex,sn-basic,referee]{sn-jnl}

\usepackage{graphicx}%
\usepackage{multirow}%
\usepackage{amsmath,amssymb,amsfonts}%
\usepackage{amsthm}%
\usepackage{mathrsfs}%
\usepackage[title]{appendix}%
\usepackage{xcolor}%
\usepackage{textcomp}%
\usepackage{manyfoot}%
\usepackage{booktabs}%
\usepackage{algorithm}%
\usepackage{algorithmicx}%
\usepackage{algpseudocode}%
\usepackage{listings}%

\theoremstyle{thmstyleone}%

\theoremstyle{thmstyletwo}%

\theoremstyle{thmstylethree}%

\begin{document}

\title[Predicting droplet landing in angle-controlled SAW jetting]{Predicting droplet landing in angle-controlled surface acoustic wave jetting}

\author[1,2]{\fnm{Md Aminul} \sur{Islam}}\email{aislam@sdsu.edu}

\author[2]{\fnm{Sara} \sur{Adibi}}\email{sadibi@sdsu.edu}

\author*[1,3]{\fnm{James} \sur{Friend}}\email{jamesfriend@wustl.edu}

\affil[1]{\orgdiv{Medically Advanced Devices Laboratory, Department of Mechanical and Aerospace Engineering}, \orgname{University of California San Diego}, \orgaddress{\street{9500 Gilman Dr.}, \city{La Jolla}, \postcode{92093}, \state{California}, \country{USA}}}

\affil[2]{\orgdiv{College of Engineering, Mechanical Engineering}, \orgname{San Diego State University}, \orgaddress{\street{5500 Campanile Drive}, \city{San Diego}, \postcode{92182}, \state{California}, \country{USA}}}

\affil*[3]{\orgdiv{Medically Advanced Devices Laboratory, Department of Mechanical Engineering and Materials Science}, \orgname{Washington University in St.\ Louis}, \orgaddress{\street{1 Brookings Drive}, \city{St.\ Louis}, \postcode{63130}, \state{Missouri}, \country{USA}}}

\abstract{Nozzle-free droplet generation eliminates the need for a dispensing orifice, but useful printing also requires control over where droplets are deposited. Here we demonstrate predictable droplet deposition using a nozzle-free, angle-controlled surface acoustic wave jetting platform. Opposing focused interdigital transducers steer droplets from a pendant parent drop toward a collection substrate, with the ejection angle set by the time-integrated acoustic acceleration imbalance. The overall acceleration ratio identifies the operating window for single-droplet ejection and separates it from no-ejection, satellite-droplet, and multiple-droplet regimes. We then connect ejection angle to landing position through a geometric deposition model referenced to a balanced shot in each experiment. At a 5.2~mm parent-drop-to-substrate spacing, 80 steered shots across twenty experiments show strongly anisotropic relative placement scatter: 0.59~mm in the steering direction and 0.06~mm transverse to it. In tests conducted independently over several days, a model calibrated on six earlier collection dates predicted landing positions on two later dates with a root-mean-square error of 0.71~mm; bootstrap resampling placed the 95\% confidence range for this error between 0.41 and 1.01~mm. The results establish a practical framework for predicting relative droplet landing in angle-controlled surface acoustic wave jetting and identify the limits imposed by steering angle, separation distance, and actuation regime from an orifice-free, clog-free droplet ejection method.}


\maketitle

\section{Introduction}\label{sec:introduction}

Precise microdroplet deposition is important for bioprinting, printed electronics, diagnostic assays, and microscale materials assembly \\citep{Truby2016,Hulme2024,advanced2025electrospray}. Conventional drop-on-demand inkjet printing has enabled scalable patterned deposition, yet the use of a nozzle can restrict the range of printable materials. Clogging, shear exposure, thermal stress, and sensitivity to particle loading are especially limiting when working with cells, soft biomaterials, polymer suspensions, and functional inks \citep{deGans2004,You2024,Zhao2022,heinrich2019,Alvarez2008}. These limitations motivate methods that can generate droplets without passing material through a small orifice prone to clogging.

Several noncontact approaches have been developed to address this challenge. Laser-induced forward transfer can deposit a wide range of materials with high spatial resolution, but often requires specialised optical components and absorbing donor layers, and jet breakup may be difficult to control for complex or viscoelastic liquids \citep{Turkoz2018,Theodorakos2019,serra2019,pique2003}. Other acoustic and cavitation-based jetting methods provide useful alternatives, although they may require complex alignment or careful control of transient regimes \citep{Roudini2023}. Acoustophoretic printing achieves material-independent drop-on-demand placement across a wide viscosity range using a Fabry--P\'erot resonator \citep{Foresti2018}, and nozzle-free acoustic droplet printing has been applied to high-concentration bioinks and tumor microenvironments \citep{KekeChen2021}; recent reviews survey the broader role of acoustics in additive manufacturing \citep{HongyiChen2025}. Within surface acoustic wave (SAW) jetting specifically, thin-film transducers have been used to vary the jetting angle over a wide range ($\sim$16--55$^\circ$) for printing applications \citep{Li2020}. Surface acoustic wave jetting offers a compact route to nozzle-free actuation because energy from a piezoelectric substrate can be coupled directly into a small parent drop to drive jet formation and droplet pinch-off \citep{Friend2011}.

SAW-based microfluidic jetting is attractive because the device is compact, the actuation is rapid, and the liquid does not need to pass through a nozzle. Prior work has shown that SAW devices can eject droplets, dispense biological samples, and support continuous or repeated droplet generation \citep{Tan2009,Bhattacharjee2011,Castro2018,connacher2018,friend2023acoustofluidics,Mei:aa,Friend2011,YeoFriend2014,Rufo2022}. Thin-film and alternative-substrate SAW devices have likewise been characterised for discrete droplet manipulation including pumping and jetting \citep{Zhou2014}, and the underlying jetting deformation and pinch-off mechanisms have been examined in both open and confined geometries \citep{LeiHu2020Review,Lei2020Actuators}. Focused interdigital transducers can localize acoustic energy at a parent drop, and opposing transducers allow the direction of the acoustic streaming impulse to be adjusted, providing a natural route to steering droplets without moving either the source drop or the printhead; numerical and experimental studies of interdigital transducer configuration have shown that transducer placement strongly influences the resulting streaming and jetting behaviour \citep{Biroun2020}.

The physics of controlled-angle SAW jetting was established by Connacher \emph{et al.} \citep{connacher2020}. That work showed that droplet ejection is not limited to the Rayleigh angle when surface tension and gravity are included with acoustic streaming, introduced a force-balance model for the ejection angle and an acceleration ratio $\Upsilon$ for identifying ejection regimes, and demonstrated how spatiotemporal actuation asymmetry can steer a droplet; subsequent reviews survey the broader SAW jetting landscape \citep{Ning2023,connacher2018}. The remaining printing questions are different: once a droplet is steered, where does it land; how large is the placement error; and under what conditions can sequential droplets form a reliable line?

This work addresses these questions. We use a downward-facing SAW jetting configuration in which droplets are ejected from a pendant parent drop toward a collection substrate. For pendant-drop SAW jetting, we translate the controlled-angle ejection model into a quantitative deposition characterization: we connect ejection angle to landing position relative to a per-experiment balanced reference, quantify the lateral and transverse placement repeatability and its anisotropy, evaluate the geometric amplification of error with angle and separation distance, test whether the ratio-to-landing mapping transfers across experiments conducted on different days, and use centerline and width metrics to characterize the one-dimensional line geometry that our method produces. We emphasize that all placement metrics here are referenced to an internal per-experiment origin rather than an externally defined substrate coordinate; the work therefore demonstrates relative landing repeatability and calibrated relative prediction, not absolute target placement. Placement precision at the sub-millimeter scale reported here is modest relative to the tens-of-micron placement achieved by established inkjet, acoustophoretic, and machine-learning-guided drop-on-demand printing \citep{Foresti2018,Derby2010,Khalil2024}; the contribution is the characterization framework and the demonstration that this clog-resistant method via SAW pendant-drop landing is repeatable and predictable in a relative sense, providing a foundation for future absolute-placement work.

\section{Results and discussion}\label{sec:results}

\subsection{Deposition framework and operating regimes}\label{subsec:framework}

\begin{figure}[h]
\centering
\includegraphics[width=\textwidth]{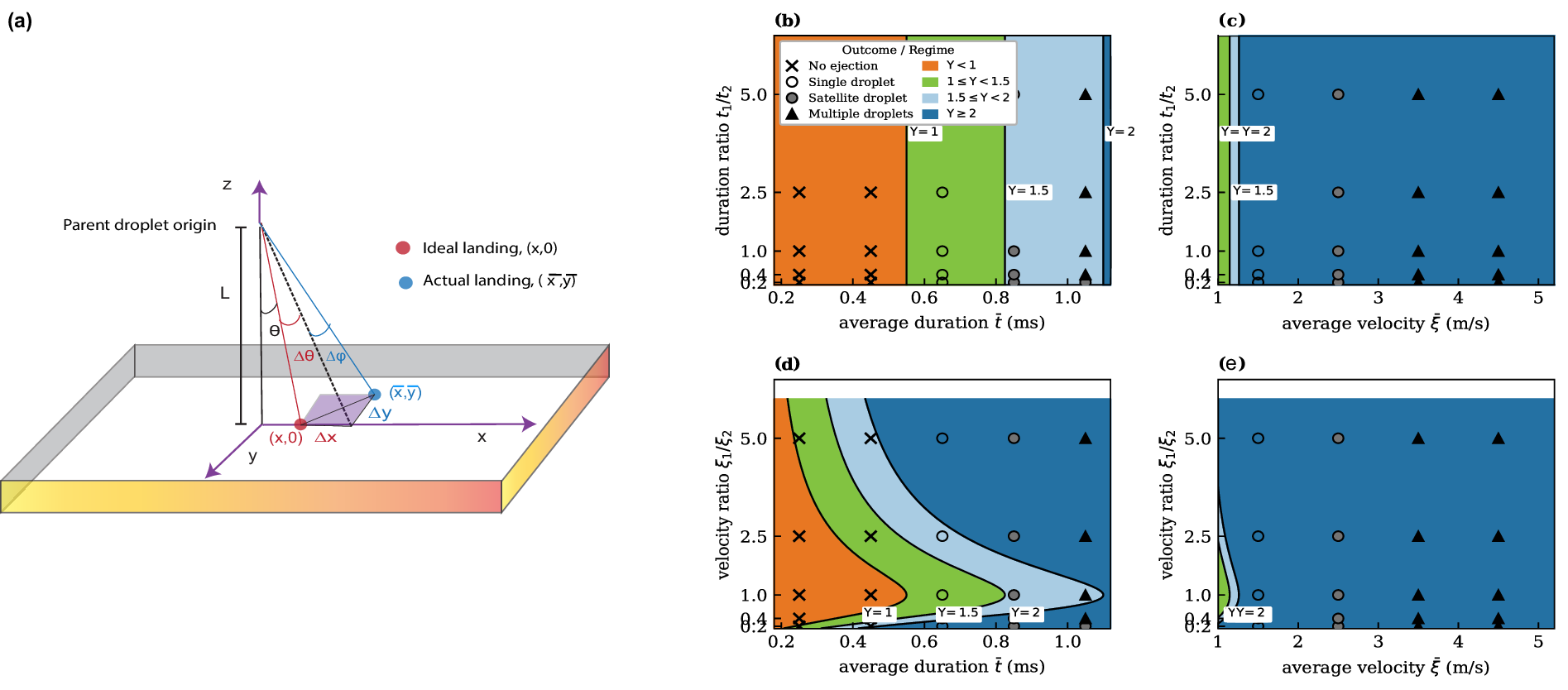}
\caption{Framework for predicting droplet landing and selecting actuation regimes. (a) Geometric definition of deposition error: a droplet ejected at angle $\theta$ has an ideal landing location $(x,0)$ after crossing separation distance $L$; in-plane angular perturbation $\Delta\theta$ and transverse perturbation $\Delta\phi$ move the measured landing position to $(\bar{x},\bar{y})$, giving errors $\Delta x$ and $\Delta y$. (b) Regime map: duration ratio $t_1/t_2$ versus average duration $\bar{t}$. (c) Duration ratio versus average vibrational velocity $\bar{\xi}$. (d) Velocity ratio $\xi_1/\xi_2$ versus average duration (corrected $\Upsilon$ contours; boundaries curve because $A_s\propto\xi^3$ makes total streaming input depend on the velocity ratio). (e) Velocity ratio versus average vibrational velocity. Colored regions denote no-ejection (orange, $\Upsilon < 1$), single-droplet (green, $1 \leq \Upsilon < 1.5$), satellite (light blue, $1.5 \leq \Upsilon < 2$), and multiple-droplet (blue, $\Upsilon \geq 2$) regimes. The single-droplet region defines the useful operating window for predictable deposition}\label{fig:fig2}
\end{figure}

Figure~\ref{fig:fig2} links the ejection model to the deposition problem. Panel~(a) shows the geometric framework: a droplet ejected at angle $\theta$ from a pendant parent drop crosses separation distance $L$ to reach the collection substrate. In-plane angular perturbation $\Delta\theta$ and transverse perturbation $\Delta\phi$ move the actual landing position away from the ideal location, giving errors $\Delta x = \bar{x} - x$ and $\Delta y = \bar{y}$. Panels~(b)--(e) show the ejection regime maps as functions of the four principal actuation parameters.

The regime maps show that steering and droplet formation are governed by different combinations of actuation parameters. The ratio quantities, $t_1/t_2$ and $v_{a,1}/v_{a,2}$, determine the signed acoustic imbalance and therefore the ejection direction. The average quantities, $\bar{t}$ and $\bar{\dot{\xi}}$, determine the total acoustic input and therefore whether a droplet is ejected cleanly. Conditions just above $\Upsilon = 1$ are most useful for deposition because they provide sufficient acoustic input for ejection without introducing satellites or multiple droplets. The boundaries $\Upsilon = 1$, $1.5$, and $2$ are empirical thresholds calibrated to the transition labels observed in the present dataset; the Connacher \emph{et al.}\ model predicts the overall scaling of $\Upsilon$ with input parameters, while the specific threshold values are adjusted here to match the experimental classifications. Operationally: a `single droplet' denotes one droplet detached cleanly per burst; a `satellite droplet' denotes one primary droplet accompanied by one or more secondary droplets smaller than half the primary diameter; `multiple droplets' denotes two or more droplets of comparable size per burst.

Because $A_s \propto v_a^3$, the total streaming from two transducers at ratio $r$ is $A_{s,1}+A_{s,2} \propto 8\bar{\xi}^3(r^3+1)/(r+1)^3$. The dimensionless acceleration ratio therefore takes the form
\begin{equation}
\Upsilon = K\,\bar{t}\,\bar{\xi}^3\,\frac{8(r^3+1)}{(r+1)^3},
\label{eq:upsilon_corrected}
\end{equation}
where $K$ absorbs the fluid and drop-volume constants. The $\Upsilon$ contours in the $(\bar{\xi},r)$ plane are curved lines that shift leftward as $r$ departs from unity because asymmetric actuation concentrates streaming into one transducer and raises the total acoustic input at the same mean velocity. At $r = 0.2$ the $\Upsilon = 1$ ejection threshold shifts to approximately $25\%$ below its symmetric value, while the effect is modest (${\sim}10\%$) near $r = 0.5$. The duration-ratio maps in panels~(b) and~(c) have straight vertical boundaries because the duration ratio does not enter the cubic velocity term; the correction applies only to panels~(d) and~(e).

\subsection{Ejection angle versus velocity ratio}\label{subsec:angle}

\begin{figure}[h]
\centering
\includegraphics[width=0.85\textwidth]{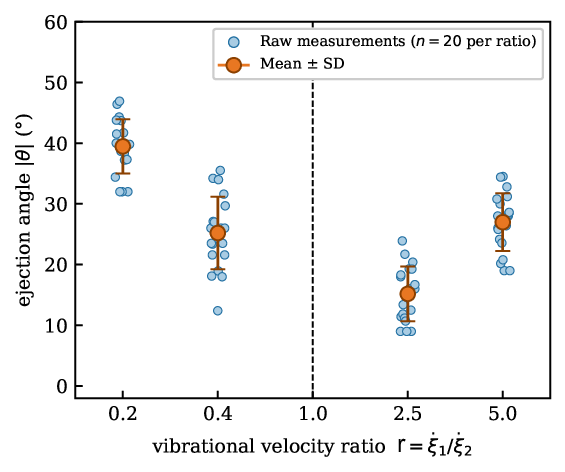}
\caption{Ejection angle $|\theta|$ as a function of vibrational velocity ratio $r = \dot{\xi}_1/\dot{\xi}_2$. Each point is one shot from a distinct experiment ($n = 20$ shots per ratio, one per experiment across 20 experiments on up to 8 collection dates; scatter therefore reflects combined shot-to-shot and day-to-day variability, not purely instantaneous repeatability). Individual measurements (light points) and mean $\pm$ standard deviation (SD; filled markers) are shown. The dashed vertical line marks $r = 1$ ($\theta = 0^\circ$). Increasing imbalance increases $|\theta|$, consistent with the signed time-integrated acceleration imbalance $t_1 A_{s,1} - t_2 A_{s,2}$}\label{fig:fig3}
\end{figure}

Figure~\ref{fig:fig3} shows the ejection angle versus vibrational velocity ratio for $n = 20$ shots per ratio (one shot per experiment across 20 experiments on up to 8 collection dates). The balanced condition $r = 1$ produces ejection near vertical ($\theta \approx 0^\circ$). As $r$ departs from unity, the acoustic imbalance grows and $|\theta|$ increases. The scatter at each ratio reflects combined shot-to-shot variability and day-to-day drift; $\sigma_\theta \approx 4.9^\circ$ estimated from this dataset is therefore an upper bound on instantaneous angular jitter.

At balanced actuation the two acoustic impulses are equal and the ejection angle approaches zero. For ratios below or above one the acoustic balance is broken and the droplet is steered away from vertical. The response is strong because $A_{s,i}$ scales with the cube of acoustic particle velocity. The effect of parent drop volume on ejection angle is discussed in Fig.~S2 of the Supplementary Information as an exploratory result; briefly, larger parent drops produce smaller ejection angles under otherwise identical actuation, and source-volume drift is therefore a candidate contributor to the day-to-day variation visible in Fig.~\ref{fig:fig3}.

From Eq.~(\ref{eq:upsilon_corrected}), the factor $8(r^3+1)/(r+1)^3$ is symmetric under $r \to 1/r$, so the ejection model predicts equal $|\theta|$ for reciprocal ratio pairs at the same mean velocity. The data show a consistent asymmetry: $\bar{|\theta|} = 39.5^\circ$ at $r = 0.2$ versus $27.0^\circ$ at $r = 5.0$, and $25.2^\circ$ at $r = 0.4$ versus $15.2^\circ$ at $r = 2.5$ (Table~S3). This systematic offset is consistent with a fixed inter-FIDT power or impedance imbalance, which is a device-level bias rather than shot-to-shot noise, that shifts the acoustic null away from the geometric center of the parent drop. Such a bias produces a constant additive offset in the signed imbalance $t_1 A_{s,1} - t_2 A_{s,2}$ and would be removed by an in-situ recalibration of the velocity ratio scale. It does not affect the deposition repeatability statistics, which are referenced to each condition's own mean landing position.

\subsection{Landing variability and angular-noise scaling}\label{subsec:repeatability}

\begin{figure}[h]
\centering
\includegraphics[width=\textwidth]{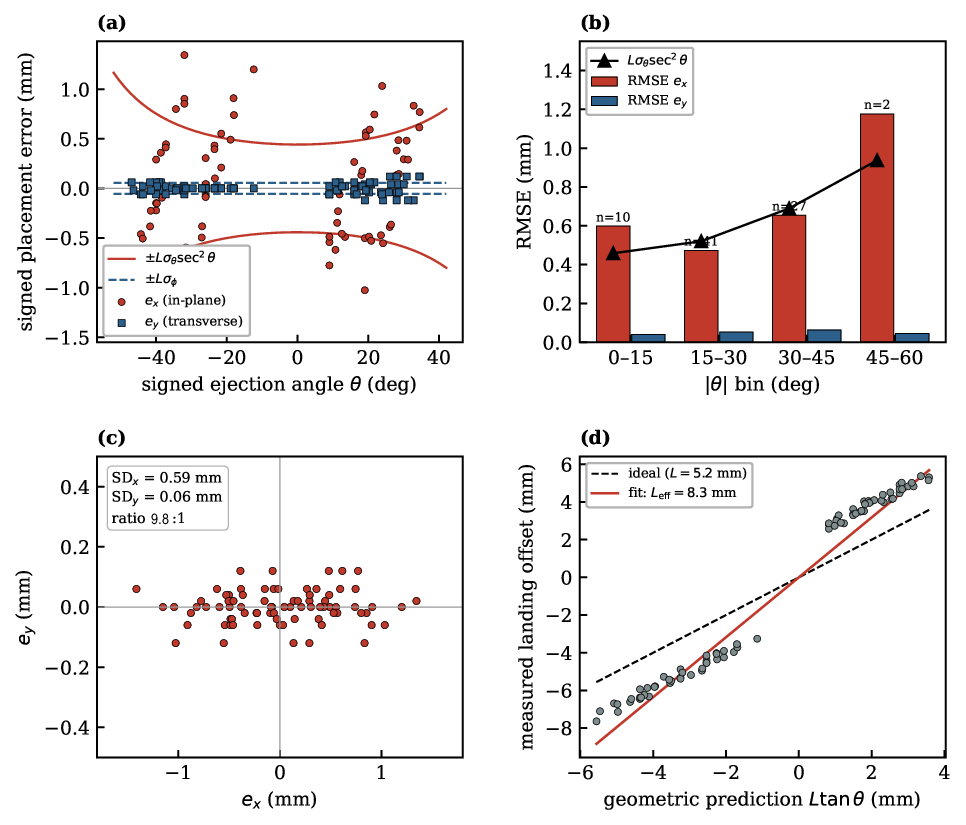}
\caption{Repeated-shot/run-to-run relative landing variability and angular-noise scaling for angle-controlled SAW jetting at $L = 5.2$~mm ($n = 80$ steered shots, one shot per ratio per experiment, PDMS-coated substrate). Errors are deviations from in-sample condition means; mean error is zero by construction. (a) Signed in-plane scatter $e_x$ and transverse scatter $e_y$ versus ejection angle $\theta$. Solid curves show the angular-noise envelope $\pm L\sigma_\theta\sec^2\theta$; dashed lines show the constant transverse envelope $\pm L\sigma_\phi$. (b) RMS scatter within absolute angle bins; the angular jitter model with $\sigma_\theta \approx 4.9^\circ$ tracks the binned in-plane RMS. (c) Anisotropy: the in-plane scatter exceeds the transverse by roughly an order of magnitude. (d) Measured landing offset versus the geometric prediction $L\tan\theta$. The slope corresponds to an empirical lever arm $L_{\mathrm{eff}} \approx 8.3$~mm, about $1.6$ times the physical separation}\label{fig:fig4}
\end{figure}

Figure~\ref{fig:fig4} characterizes shot-to-shot landing placement variability at $L = 5.2$~mm across twenty experiments (one shot per ratio per experiment). Within each experiment the $r = 1$ landing defines the coordinate origin $(x,y) = (0,0)$; the signed errors $e_x$ and $e_y$ are therefore deviations from each commanded condition's in-sample mean, and the overall mean error is zero by construction. This analysis quantifies angular-noise scaling and placement variability rather than independent target prediction, which is addressed separately in Sect.~\ref{subsec:predictive}. Panel~(a) shows the signed errors versus ejection angle with the angular noise envelopes $\pm L\sigma_\theta\sec^2\theta$ and $\pm L\sigma_\phi$. Panel~(b) reports the root mean square error within absolute angle bins. Panel~(c) shows the error, and panel~(d) compares the measured landing against the geometric prediction $L\tan\theta$.

\begin{table}[h]
\caption{Signed variability statistics about in-sample condition means for the $n = 80$ steered shots at $L = 5.2$~mm (PDMS substrate, one shot per ratio per experiment). Errors are deviations from the per-condition mean, so the overall mean and bias are zero by construction; these values characterize angular-noise scaling and anisotropy, not independent prediction accuracy. The final column is the 95\% CI half-width of the condition mean}\label{tab:fig5_statistics}
\begingroup
\setlength{\tabcolsep}{4pt}
\renewcommand{\arraystretch}{1.15}
\begin{tabular*}{\linewidth}
  {@{\extracolsep{\fill}}l*{7}{c}@{}}
\toprule
Component & $n$ &
\shortstack{Mean\\(mm)} &
\shortstack{SD\\(mm)} &
\shortstack{MAE\\(mm)} &
\shortstack{RMSE\\(mm)} &
\shortstack{Bias\\(mm)} &
\shortstack{95\% CI half-width\\(mm)} \\
\midrule
$e_x$ & 80 & 0.00 & 0.59 & 0.48 & 0.58 & 0.00 & 0.13 \\
$e_y$ & 80 & 0.00 & 0.06 & 0.04 & 0.06 & 0.00 & 0.01 \\
\midrule
\multicolumn{8}{@{}p{\linewidth}@{}}{%
  \footnotesize
  MAE, mean absolute error; RMSE, root mean square error;
  CI, confidence interval. The 95\% CI is reported
  as its half-width.
} \\
\botrule
\end{tabular*}
\endgroup
\end{table}

Table~\ref{tab:fig5_statistics} quantifies the anisotropy. The in-plane error has a standard deviation of $0.59$~mm and grows with steering angle, whereas the transverse error has a standard deviation of $0.06$~mm --- a $\sim 10{:}1$ ratio (precisely $9.8{:}1$). The angular jitter model $L\sigma_\theta\sec^2\theta$, with $\sigma_\theta \approx 4.9^\circ$ calibrated from the same dataset, tracks the measured in-plane RMSE across the populated angle bins. Because $\sigma_\theta$ is estimated in-sample, this comparison is a self-consistency check confirming that the in-plane scatter scales as $\sec^2\theta$, not an out-of-sample prediction. The $45$--$60^\circ$ bin ($n = 2$) is excluded from this quantitative comparison. The transverse error remains bounded because it depends only on the smaller out-of-plane perturbation $\Delta\phi$. These results support the following variability statement: run-to-run relative placement variability is strongly anisotropic. The in-plane component, consistent with angular jitter amplified by $\sec^2\theta$, grows from $\sim\!0.5$~mm near $\theta = 0^\circ$ and approaches $1$~mm in the sparsely sampled $45$--$60^\circ$ bin ($n = 2$). The transverse component remains below $0.1$~mm across all conditions. Independent predictive accuracy is evaluated next in Sect.~\ref{subsec:predictive}.

High-speed imaging provides measurements of the exit velocity $U$ and neck diameter $d_{\mathrm{neck}}$ ($n = 1$ representative frame per condition). The measured values span $U = 0.25$~m/s at $r = 0.2$ ($\theta \approx 45^\circ$) to $U = 1.20$~m/s at $r = 1.0$ ($\theta = 0^\circ$), and $d_{\mathrm{neck}}$ increases from approximately $50~\mu$m to $160~\mu$m over the same range. Applying the Rayleigh breakup scaling \citep{Eggers2008} $d_{\mathrm{drop}} \approx 1.9\,d_{\mathrm{neck}}$ gives ejected droplet diameters from approximately $95~\mu$m at $r = 0.2$ to $300~\mu$m at $r = 1.0$. The in-plane variability standard deviation of $\sigma_x = 0.59$~mm corresponds to $6.2\,d_{\mathrm{drop}}$ at the most strongly steered condition ($r = 0.2$, $d_{\mathrm{drop}} \approx 95~\mu$m) and $2.0\,d_{\mathrm{drop}}$ at $r = 5.0$ ($d_{\mathrm{drop}} \approx 300~\mu$m), showing that placement scatter relative to droplet size decreases as the velocity ratio approaches unity, because less-steered droplets are larger and experience lower $\sec^2\theta$ amplification.

At the measured $L = 5.2$~mm, the best-fit slope of measured landing against $L\tan\theta$ corresponds to an empirical lever arm $L_{\mathrm{eff}} \approx 8.3$~mm (Fig.~\ref{fig:fig4}d), i.e.\ $L_{\mathrm{eff}}/L \approx 1.6$, a $\sim 60\%$ larger effective lever arm than the physical separation. (The slope is reported with a near-zero intercept; an alternative fit allowing a non-zero intercept reduces the apparent slope. Moreover, the per-condition ratio $\hat{x}(r)/[L\tan\bar{\theta}]$ varies from $\sim\!1.4$ to $\sim\!2.3$ across the four ratios, so $L_{\mathrm{eff}}$ is a single-parameter summary of a structured residual rather than a true physical length, and the noise envelope in panel~(a) is computed at the physical $L = 5.2$~mm.) A ballistic estimate shows that at $r = 0.2$ ($\theta \approx 45^\circ$, $U = 0.25$~m/s) the straight projection gives $x = 5.2$~mm while the ballistic result ($t_f = (-U\cos\theta + \sqrt{U^2\cos^2\theta + 2gL})/g$, $x_{\mathrm{bal}} = U\sin\theta\,t_f$) gives $x_{\mathrm{bal}} \approx 3.4$~mm---an inward shift, not an outward one. Ballistic curvature therefore cannot explain the overshoot; it acts in the opposite direction. The discrepancy is attributed to a combination of: trajectory-angle measurement rather than immediately before impact, systematic offset in the per-experiment $r = 1$ coordinate origin, and post-impact centroid displacement \citep{Yarin2006,Josserand2016} on the PDMS-coated substrate. Without fixed-reference absolute position measurements, these three contributions cannot be separated from the available data; $L_{\mathrm{eff}}$ is therefore reported as a descriptive empirical observation. Future work measuring the $r = 1$ landing against a fixed geometric target will separate origin-placement bias from spreading and trajectory effects. Reducing $L$ toward the millimeter scale, as in a compact or handheld printhead \citep{Huang2021}, is expected to reduce all three contributions and is identified as a primary route to improved placement accuracy.

\subsection{Held-out predictive deposition test}\label{subsec:predictive}

\begin{figure}[h]
\centering
\includegraphics[width=0.85\textwidth]{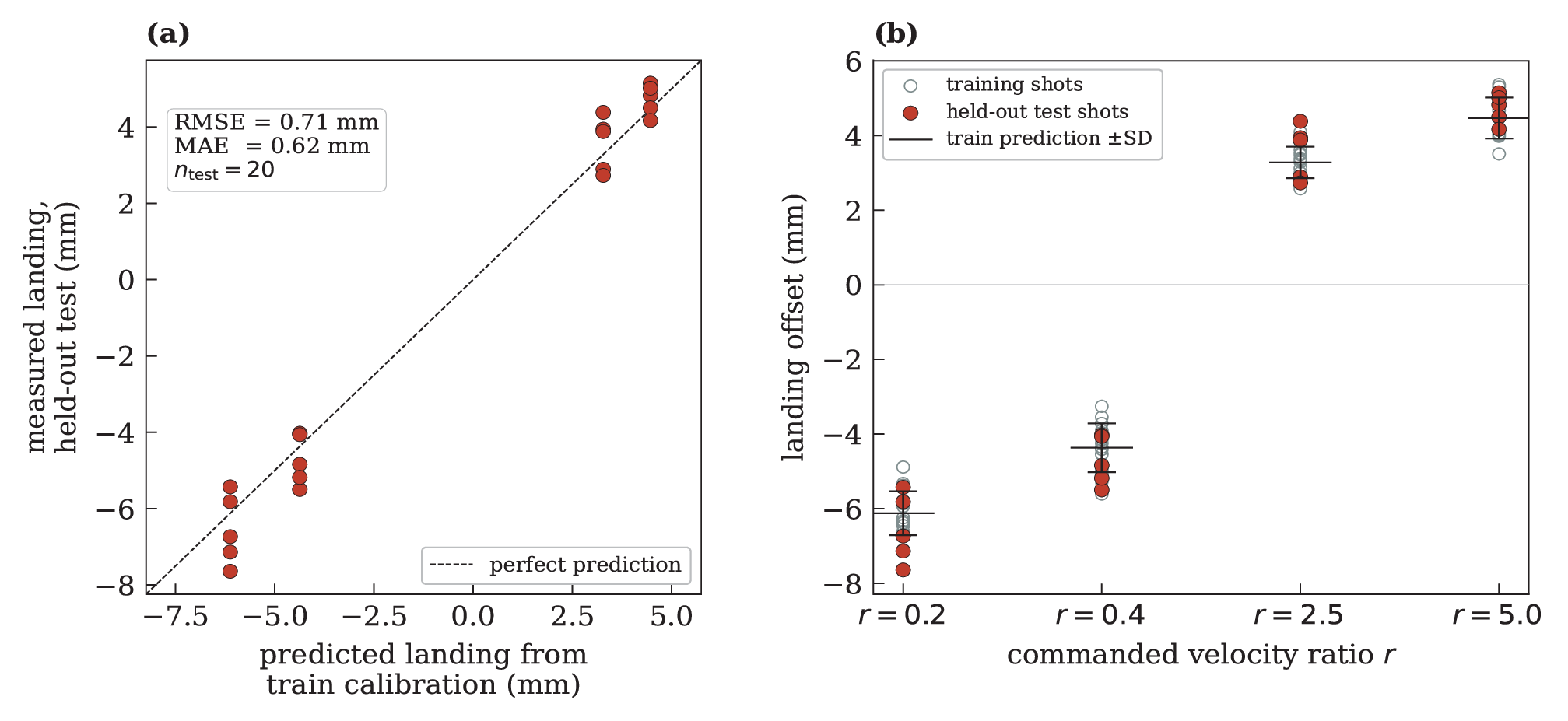}
\caption{Held-out predictive deposition test ($n_{\mathrm{train}}=60$, dates 1--6; $n_{\mathrm{test}}=20$, dates 7--8; one shot per ratio per experiment). The per-experiment $r=1$ reference shot defines the coordinate origin independently on training and test days. The training-day condition mean at each ratio is used as the predicted landing for test-day shots. (a) Parity plot: predicted versus measured in-plane landing on the 20 held-out shots (RMSE $= 0.71$~mm, MAE $= 0.62$~mm). (b) Per-ratio landing positions for training shots (open circles) and held-out test shots (filled circles) with the training-day mean and standard deviation; the mapping transfers to the later dates, with the largest residual drift at $r = 0.2$, the strongest steering condition}\label{fig:fig5}
\end{figure}

Figure~\ref{fig:fig5} presents the held-out predictive test. The dataset spans eight different collection dates (20 experiments total; one shot per ratio per experiment, four steered ratios $r \in \{0.2, 0.4, 2.5, 5.0\}$ plus one reference shot at $r = 1$). The per-experiment $r = 1$ shot defines the coordinate origin independently for each test-day experiment, so no training-day absolute position information is used on test days. The commanded-ratio-to-landing mapping---the mean in-plane landing position at each ratio---was estimated from the six earliest collection dates ($n_{\mathrm{train}} = 60$; 15 shots per ratio on dates 1--6, one shot per ratio per experiment across 15 experiments) and applied without recalibration to the two later dates ($n_{\mathrm{test}} = 20$; 5 shots per ratio on dates 7--8, one shot per ratio per experiment across 5 experiments).

The held-out prediction achieved RMSE $= 0.71$~mm and MAE $= 0.62$~mm. Given the test-set size ($n_{\mathrm{test}} = 20$ shots across 5 experiments), the bootstrap 95\% confidence interval on the RMSE is approximately $0.41$--$1.01$~mm; this uncertainty should be borne in mind when interpreting the precision of the result. The largest residual errors occur at $r = 0.2$, the strongest steering condition, where day-to-day drift is most amplified by the $\sec^2\theta$ projection. The four-ratio mapping transferred to the later experimental days, demonstrating calibrated relative prediction at this scale. We note two limitations of this test. First, the predictor is a four-value lookup of the training-day mean at each commanded ratio; it does not interpolate to untested ratios and does not use the physics model, so it should be read as evidence that the ratio-to-landing relationship is stable across days rather than as a general predictive model. Second, the baseline against which it is compared (predicting the global mean, which is zero by the per-experiment origin convention) is weak, so the apparent improvement factor mainly reflects that different ratios land in systematically different places. Full per-ratio held-out errors, training details, and the baseline comparison are provided in Sect.~S10 of the Supplementary Information.

\subsection{Illustrative line deposition}\label{subsec:line}

\begin{figure}[h]
\centering
\includegraphics[width=\textwidth]{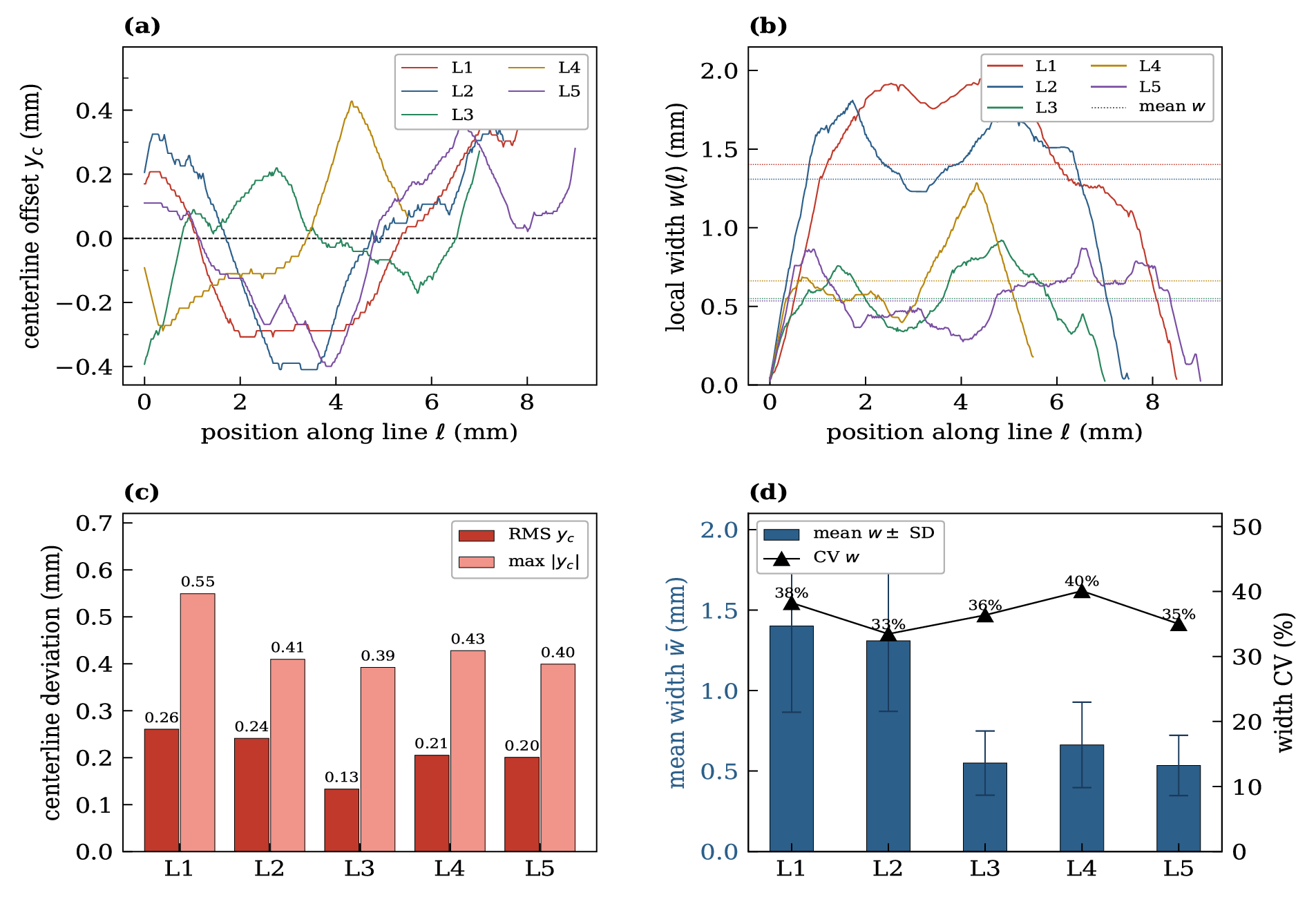}
\caption{Quantitative assessment of one-dimensional line deposition fidelity for five printed lines (all quantities in mm). (a) Centerline offset $y_c(\ell)$ along each line; the dashed line marks the ideal $y_c = 0$. This measures local straightness about each line's own mean path, not absolute placement relative to a commanded target. (b) Local width $w(\ell)$ along each line; dotted lines mark each line's mean width. (c) Summary centerline deviation per line: root mean square $y_c$ and maximum $|y_c|$. (d) Summary width statistics per line: mean width with standard deviation (bars) and coefficient of variation (markers). The five lines differ in length and width and represent different operating conditions rather than replicates; the hydrophilic-treated surface introduces wetting and coalescence as confounding variables relative to the single-droplet experiments}\label{fig:fig6}
\end{figure}

After characterizing single-droplet repeatability and calibrated relative prediction, we examined the internal geometry of one-dimensional patterns formed by sequential deposition. This analysis characterizes the \emph{internal geometric fidelity} of each deposited line, that is, its straightness relative to its own mean path and the uniformity of its local width, rather than its placement relative to an externally commanded target, which the present substrate does not reference. Figure~\ref{fig:fig6} shows the extracted line metrics for five printed lines using two independent quantities: the centerline offset $y_c(\ell) = y_{\mathrm{centerline}}(\ell) - \overline{y_{\mathrm{centerline}}}$, which measures local straightness about each line's own mean path, and the local width $w(\ell)$, measured perpendicular to the skeleton centerline \citep{Soltman2008,Derby2010,Stringer2010,Lohse2022}.

The five lines used the same device and general actuation range as the single-droplet experiments, but each line was printed under a different command condition; the five lines are therefore not statistical replicates. The substrate was coated with a hydrophilic solution to obtain a contact angle of $\approx 7.6^\circ$ to promote lateral spreading and coalescence. The number of drops deposited per line was as follows: line~1 (12 drops), line~2 (10 drops), line~3 (8 drops), line~4 (11 drops), line~5 (13 drops). Spreading on the hydrophilic surface is the dominant source of width variability and is a confound relative to the steering variability measured in the single-droplet experiments. Segmented masks with centerlines are provided in Fig.~S4 of the Supplementary Information.

\begin{table}[h]
\caption{Line fidelity metrics extracted from the centerline and local width profiles in Fig.~\ref{fig:fig6} (all values in mm). Centerline straightness is quantified by the root mean square offset $y_{\mathrm{RMS}}$, mean absolute offset, and maximum absolute offset about each line's own mean path. Width uniformity is quantified by the mean local width $\bar{w}$, its standard deviation, and its coefficient of variation. The five lines are different operating conditions, not replicates; the CV range therefore describes individual outcomes, not a statistical characterization of line-fidelity precision}\label{tab:line_fidelity}
\begin{tabular*}{\textwidth}{@{\extracolsep\fill}lcccccc}
\toprule
Line & $y_{\mathrm{RMS}}$ & Mean $|y_c|$ & $|y_c|_{\max}$ & $\bar{w}$ & SD $w$ & $CV_w$ (\%) \\
\midrule
Line 1 & 0.260 & 0.232 & 0.549 & 1.402 & 0.536 & 38.3 \\
Line 2 & 0.241 & 0.204 & 0.410 & 1.308 & 0.438 & 33.5 \\
Line 3 & 0.133 & 0.105 & 0.392 & 0.549 & 0.200 & 36.4 \\
Line 4 & 0.205 & 0.178 & 0.428 & 0.662 & 0.265 & 40.1 \\
Line 5 & 0.201 & 0.173 & 0.399 & 0.535 & 0.187 & 35.0 \\
\botrule
\end{tabular*}
\end{table}

Among the five lines, line~3 had the smallest centerline root mean square offset ($0.13$~mm) and line~2 had the lowest width coefficient of variation ($33.5\%$). Because each line was printed under a different command condition with $n = 1$ per condition, and because spreading on the hydrophilic surface is the dominant width confound, these values are reported as a qualitative proof-of-concept demonstration that continuous one-dimensional deposition with quantifiable internal geometry is achievable --- not as a statistical characterization of line-fidelity precision. A systematic study would require replicate lines at a single fixed command condition; this is identified as future work. The demonstration nonetheless shows that both centerline and width metrics are needed to describe line geometry: a line can remain near its own mean path while showing local width variability, or hold uniform width while its centerline drifts.

\subsection{Estimated recovery-limited repetition rate}\label{subsec:rate}

Printing rate was estimated from high-speed images. For each input condition, the ejection onset time $t_i$ and the end of visible relaxation $t_f$ were identified frame by frame. The recovery-limited cycle frequency was estimated as
\begin{equation}
f_{\mathrm{est}} = \frac{1}{t_f - t_i}.
\label{eq:print_rate}
\end{equation}
Across $n = 5$ analysed ejection intervals, the mean estimated rate was 7.90~Hz and the highest single-interval value was 8.83~Hz. These values represent inverse relaxation-interval estimates obtained under manual triggering, not demonstrated sustained throughput. Automated waveform timing and closed-loop triggering are expected to improve both rate stability and deposition repeatability.

\section{Conclusions}\label{sec:conclusions}

This work reframes angle-controlled SAW jetting as a deposition control problem. The ejection angle model and regime parameter from Connacher \emph{et al.}\ provide the basis for steering and for selecting single-droplet conditions. We extend that framework by linking ejection angle to landing position, measuring deposition errors, and quantifying line formation. The signed time-integrated acceleration imbalance $t_1 A_{s,1} - t_2 A_{s,2}$ controls the direction of ejection, while $\Upsilon$ identifies the ejection regime.

The landing model explains why deposition error is anisotropic: in-plane error grows as $L\Delta\theta\sec^2\theta$ and is therefore amplified at high steering angles and large separation distances, while transverse error depends primarily on the smaller perturbation $L\Delta\phi$. Twenty experiments at $L = 5.2$~mm characterized shot-to-shot repeatability: in-plane scatter SD $= 0.59$~mm, transverse scatter SD $= 0.06$~mm (a $\sim 10{:}1$ ratio), with the in-plane scatter consistent with angular jitter of $\sigma_\theta \approx 4.9^\circ$ amplified by the $\sec^2\theta$ projection factor. A separate held-out date-split test, calibrating the four-ratio landing mapping on six earlier collection dates and predicting on two later dates, achieved RMSE $= 0.71$~mm (bootstrap 95\% CI $0.41$--$1.01$~mm; $n_{\mathrm{test}} = 20$ shots across 5 experiments), a first demonstration of calibrated relative landing prediction for this platform. All placement metrics are referenced to a per-experiment balanced shot; the work therefore establishes run-to-run relative placement variability and placement prediction, while absolute target placement remains for future work. The deposited-centroid span exceeds the straight-line projection at $L = 5.2$~mm; ballistic curvature acts in the opposite direction to the observed overshoot, and the discrepancy is attributed to trajectory-angle measurement, per-experiment origin placement, and post-impact centroid displacement on the PDMS substrate. Reducing $L$ toward the millimeter scale, as in a compact or handheld printhead \citep{Huang2021}, is expected to suppress these contributions and is identified as a primary route to improved placement accuracy.

A proof-of-concept line-deposition study showed that the internal geometry of one-dimensional patterns, namely centerline straightness and local width uniformity, can be quantified, with width coefficients of variation of $33$--$40\%$ across five single-condition demonstrations; a statistically powered line-fidelity study at fixed conditions remains future work. The kinematic projection from a known exit direction to a landing point is independent of fluid properties under the straight-trajectory approximation; the ejection window, angular scatter, droplet breakup, exit speed, and post-impact spreading remain fluid dependent. Specific near-term steps include: (i) reducing $L$ below 2~mm to test the predicted suppression of $L_{\mathrm{eff}}$ overshoot; (ii) volume-controlled dispensing to separate parent-drop-volume drift from actuation variability (see Sect.~S4 of the Supplementary Information); and (iii) replicate line experiments at a fixed command condition for a rigorous statistical characterization of line fidelity. Future work will also address automated waveform control, real-time feedback, and extended material sets including bio-inks and functional inks \citep{Rufo2022,Jentsch2021}.

\section{Materials and methods}\label{sec:methods}

\subsection{SAW device assembly and experimental configuration}\label{subsec:device}

SAW devices were fabricated on $127.68^\circ$ Y-rotated, X-propagating lithium niobate (LiNbO$_3$) substrates (Roditi Ltd., UK) and were designed to operate near 40~MHz. Focused interdigital transducers were patterned using photolithography and electron-beam metal deposition as described by \citet{Mei:aa}. Each focused interdigital transducer (FIDT) consisted of 20 finger pairs arranged in a mirrored arc geometry so that acoustic energy was focused at the device center.

A Luer-lock syringe was mounted vertically above the acoustic focal region using a rigid holder. Deionised water was used as the working liquid for the main experiments. Pendant parent drops were formed at the syringe tip and allowed to stabilise before actuation. The collection substrate was positioned below the pendant parent drop on a manually adjustable vertical $z$ stage. The separation distance, $L$, was defined as the vertical distance between the droplet origin and the collection substrate. Glass microscope slides were used as collection substrates. For single-droplet deposition experiments, slides were coated with polydimethylsiloxane (PDMS, Sylgard 184, Dow Corning, 10:1 base to curing agent ratio, cured at $60^\circ$C for 24~h). For line deposition experiments, substrates were treated with a commercial hydrophilic coating (Aculon, San Diego, CA, USA) to promote spreading and continuous line formation; this treatment produces a static contact angle of approximately $7.6^\circ$, near the conventional superhydrophilic threshold ($\theta < 10^\circ$), and promotes rapid lateral spreading upon droplet impact.

\begin{figure}[h]
\centering
\includegraphics[width=\textwidth]{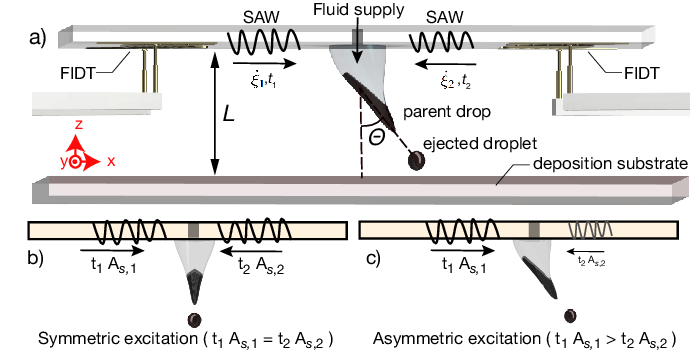}
\caption{Experimental configuration for angle-controlled SAW droplet deposition. A pendant parent drop is held below the syringe and above a collection substrate separated by distance $L$. Opposing focused interdigital transducers generate acoustic streaming toward the parent drop. When the acoustic impulses are balanced, $t_1 A_{s,1} = t_2 A_{s,2}$, the droplet ejects vertically with $\theta = 0^\circ$. When the impulses are unequal, $t_1 A_{s,1} \neq t_2 A_{s,2}$, the droplet ejects at a signed angle $\theta$ toward the weaker side}\label{fig:fig1}
\end{figure}

\subsection{Acoustic actuation, calibration, and imaging}\label{subsec:actuation}

The two FIDTs were driven using a dual-channel function generator (WF1968, NF Corporation, Japan) and broadband RF amplifiers (Model 5U1000, Amplifier Research). Sinusoidal bursts were applied to FIDT1 and FIDT2 with independently assigned burst durations $t_1$ and $t_2$. Actuation asymmetry was introduced either by changing the vibrational velocity ratio, $\dot{\xi}_1/\dot{\xi}_2$, or by changing the burst duration ratio, $t_1/t_2$. The average burst duration and average vibrational velocity were defined as
\begin{equation}
\bar{t}=\frac{t_1+t_2}{2},
\qquad
\bar{\dot{\xi}}=\frac{\dot{\xi}_1+\dot{\xi}_2}{2}.
\label{eq:averages}
\end{equation}

The focal vibrational velocity was measured by laser Doppler vibrometry (LDV; UHF 120, Polytec, Germany); LDV calibration curves for both transducers are provided in Fig.~S1 of the Supplementary Information. The measured focal velocity was used to relate the applied input voltage to the acoustic velocity used in the model, following the calibration approach of \citet{connacher2020}. High-speed imaging was performed at 4000~fps using a Fastcam Mini UX100 camera (Photron) with a Nikon 105~mm f/2.8 Micro NIKKOR lens. A diffuse LED source provided back-illumination. Image sequences were analysed in ImageJ and MATLAB to extract ejection angle, droplet trajectory, and line profiles. Unless otherwise stated, repeated measurements are reported as mean values with one standard deviation.

Landing position was determined from top-view images of the dried deposit recorded after each deposited droplet had fully evaporated on the substrate. The centroid of the dried deposit was located in each image after background subtraction using a thresholded binary mask; centroid coordinates were extracted in ImageJ and converted to physical units using the pixel-to-length calibration described in Sect.~S7 of the Supplementary Information. Because the droplet was allowed to dry completely before imaging, the measured centroid reflects the final deposit position rather than a transient spreading state. On the PDMS-coated substrate, evaporation of deionised water produces a thin deposit ring whose centroid was used as the landing position; any systematic offset between this dried-deposit centroid and the true impact centroid is expected to be small and approximately symmetric for the conditions studied, and is subsumed into the empirical lever arm $L_{\mathrm{eff}}$ discussed in Sect.~\ref{subsec:repeatability}. The balanced $r = 1$ deposit centroid in each experiment served as the coordinate origin, so that all reported placement errors are relative displacements rather than absolute substrate coordinates.

\subsection{Ejection angle and regime model}\label{subsec:model}

The ejection angle model used here follows the force-balance framework of \citet{connacher2020}. We use it as the basis for controlled droplet steering, then extend the analysis to deposition error. The acoustic streaming acceleration generated by the $i^\mathrm{th}$ FIDT is
\begin{equation}
A_{s,i}
=
\frac{\pi v_{a,i}^{3} f \beta}{2c_l^2},
\label{eq:streaming_acceleration}
\end{equation}
where $v_{a,i}$ is the acoustic particle velocity associated with FIDT $i$, $f$ is the actuation frequency, $\beta$ is the nonlinear compressibility parameter of the liquid, and $c_l$ is the sound speed in the liquid. Because $A_{s,i}$ scales with $v_{a,i}^{3}$, small changes in velocity can produce large changes in acoustic input as shown by \citet{Orosco2022}.

For two opposing FIDTs, the signed ejection angle including acoustic streaming, time-averaged surface tension, and gravity is
\begin{equation}
\theta
=
\arctan
\left[
\frac{
(t_1 A_{s,1}-t_2 A_{s,2})\sin\theta_R
}{
(t_1 A_{s,1}+t_2 A_{s,2})\cos\theta_R
-
t_{\mathrm{df}}
\!\left(
D_{\mathrm{st}}\sin\theta_s/4+g
\right)
}
\right],
\label{eq:theta_full_model}
\end{equation}
where $\theta_R$ is the Rayleigh angle, $\theta_s$ is the local interface angle, $g$ is gravitational acceleration (noting that in the pendant geometry used here gravity acts in the direction of ejection, reversing its sign relative to the sessile-drop derivation of Connacher \emph{et al.}), and $D_{\mathrm{st}} = 2\pi\gamma r_s/(V\rho_l)$ is the surface tension acceleration with $\gamma$ the surface tension, $r_s$ the local radius of curvature, $V$ the parent drop volume, and $\rho_l$ the liquid density. The local interface angle $\theta_s$ is taken from high-speed images of the deforming parent drop at the moment of column formation. The deformation time
\begin{equation}
t_{\mathrm{df}}
=
\frac{a - \alpha_r^{-1}/4}{\sin(\theta_R)\,v_a}
+
\sqrt{\frac{\rho_l r_s^3}{\gamma}}
\label{eq:deformation_time}
\end{equation}
accounts for the column formation time and the capillary breakup time, where $a$ is the parent drop footprint radius and $\alpha_r^{-1}$ is the Rayleigh wave attenuation length (the characteristic distance over which the surface wave amplitude decays as it couples into the liquid).

Steering is controlled by the signed time-integrated acceleration imbalance $t_1 A_{s,1} - t_2 A_{s,2}$. Balanced actuation gives $t_1 A_{s,1} = t_2 A_{s,2}$ and predicts $\theta = 0^\circ$. Unequal actuation produces a finite signed imbalance and therefore a nonzero ejection angle. Direction can accordingly be controlled by changing $t_1/t_2$, $v_{a,1}/v_{a,2}$, or both of them.

The ejection regime was classified using the acceleration ratio
\begin{equation}
\Upsilon
=
\frac{
(t_1 A_{s,1}+t_2 A_{s,2})\cos\theta_R
}{
t_{\mathrm{df}}(D_{\mathrm{st}}/4+g)
}.
\label{eq:upsilon}
\end{equation}
This ratio compares the acoustic streaming input with the restoring effects of surface tension and gravity. Following the Connacher interpretation: $\Upsilon < 1$ indicates no ejection, values just above one produce single-droplet ejection, intermediate values produce satellite droplets, and larger values produce multiple droplets. The single-droplet region defines the useful operating window for predictable deposition.

\subsection{Landing position model and statistical measures}\label{subsec:landingmodel}

Neglecting ballistic propagation of the ejected droplets based on the short times and small distances involved, and therefore assuming a simple geometric relation instead, for an ideal trajectory in the $xz$ plane, a droplet ejected at angle $\theta$ lands at
\begin{equation}
x = L\tan\theta.
\label{eq:ideal_landing}
\end{equation}
Small angular perturbations $\Delta\theta$ (in the $xz$ plane) and $\Delta\phi$ (transverse in the $yz$ plane) shift the measured landing position $(\bar{x},\bar{y})$ away from the ideal position $(x,0)$:
\begin{align}
\bar{x} &= L\tan(\theta+\Delta\theta), \label{eq:xbar}\\
\bar{y} &= L\tan(\Delta\phi).          \label{eq:ybar}
\end{align}
For small angular errors the linearized forms are
\begin{align}
\Delta x &= L\!\left[\tan(\theta+\Delta\theta)-\tan\theta\right]
            \approx L\Delta\theta\sec^2\theta,
\label{eq:dx_linearized}\\
\Delta y &= L\tan(\Delta\phi) \approx L\Delta\phi.
\label{eq:dy_linearized}
\end{align}
In-plane error therefore grows with $\sec^2\theta$, while transverse error remains approximately proportional to $L\Delta\phi$. In both directions, the separation distance $L$ acts as a lever arm. A droplet reaches the substrate after traveling over a path set by the drop-to-substrate spacing, so a larger $L$ projects the same angular jitter over a longer distance and increases the scatter in the landing position. The two directions differ in their dependence on steering angle. The in-plane error includes the $\sec^2\theta$ amplification in Eq.~(\ref{eq:dx_linearized}), whereas the transverse error in Eq.~(\ref{eq:dy_linearized}) has no corresponding angular amplification and therefore remains nearly angle independent. This separation-distance scaling is examined in Fig.~S3 of the Supplementary Information.

Landing repeatability is quantified using signed errors $e_x = x_{\mathrm{meas}} - x_{\mathrm{pred}}$ and $e_y = y_{\mathrm{meas}} - y_{\mathrm{pred}}$, where $x_{\mathrm{pred}}$ and $y_{\mathrm{pred}}$ are the in-sample condition means for each commanded ratio. Because the reference is the mean, the mean error is zero by construction; the statistics therefore characterize shot-to-shot angular scatter rather than independent target prediction. Standard deviation, mean absolute error, root mean square error, and the 95\% confidence-interval half-width are reported.

Within each experiment, the balanced $r = 1$ shot defines the per-experiment coordinate origin: the landing position of the $r = 1$ droplet is treated as the coordinate zero for all other shots in that same experiment, so that inter-experiment variability in absolute device position does not enter the placement errors. The balanced shots are retained and reported as reference measurements but are excluded from Table~\ref{tab:fig5_statistics} because they define the run-level origin. Independent prediction accuracy is evaluated in Sect.~\ref{subsec:predictive}. The ratio-to-landing mapping is trained only on earlier collection dates and tested on later dates; each test-day experiment still uses its own balanced single ($r=1$) shot to define the relative coordinate origin.

\backmatter

\bmhead{Supplementary information}

This article has accompanying Supplementary Information containing: device and acquisition parameters (Table~S1) and fixed physical constants (Table~S2); LDV calibration curves for both transducers (Fig.~S1); ejection angle as a function of parent drop volume under fixed asymmetric actuation (Fig.~S2, exploratory, $n = 1$ per volume); height-dependent deposition error measurements at $L = 3.8$, 5.2, and 6.6~mm as a geometric scaling check (Fig.~S3); segmented line masks with extracted centerlines and local width profiles for the five printed lines (Fig.~S4); image analysis procedures (Sect.~S7); per-condition and angle-binned deposition statistics (Tables~S3 and S4); the held-out predictive test protocol with per-ratio errors (Sect.~S10, Table~S5); and full line-fidelity statistics (Table~S6).

\bmhead{Acknowledgements}
The authors thank colleagues in the Medically Advanced Devices Laboratory for helpful discussions, and J.~Orosco for advice on acoustic streaming models.

\section*{Declarations}

\bmhead{Funding}
J.~Friend is grateful to the Office of Naval Research, United States (grant 13423461), the National Science Foundation (grant ECCS 2314118), and support from UC San Diego via the Penner Endowed Chair of Engineering and Physics for support of this work.

\bmhead{Conflict of interest}
The authors declare no competing interests.

\bmhead{Ethics approval and consent to participate}
Not applicable.

\bmhead{Consent for publication}
All authors have read and approved the final manuscript and consent to its publication.

\bmhead{Data availability}
The processed datasets, calibration files, and analysis scripts that support the findings of this study are openly available in a Zenodo repository at \url{https://doi.org/10.5281/zenodo.21071811}. Raw high-speed image sequences are available from the corresponding author upon reasonable request owing to their size.

\bmhead{Materials availability}
Not applicable.

\bmhead{Code availability}
The Python analysis and plotting scripts supporting all figures and tables are included in the Zenodo repository cited under Data availability.

\bmhead{Author contribution}
M.A.I.\ performed the experiments, analysed the data, prepared the figures, and drafted the manuscript. S.A.\ contributed to experimental planning, data interpretation, and manuscript revision. J.F.\ conceived and supervised the project, contributed to model development and interpretation, and revised the manuscript. All authors reviewed and approved the final manuscript.

\bibliography{sn-bibliography}

@article{Roudini2023,
	address = {SAWLab Saxony, Leibniz Institute for Solid State and Materials Research Dresden, Helmholtzstr. 20, Dresden 01069, Germany. Electronic address: m.roudini@ifw-dresden.de.; Otto von Guerricke University, Institute for Physics, Universit{\"a}tsplatz. 2, Magdeburg 39106, Germany.; Technion-Israel Institute of Technology, Department of Chemical Engineering, Haifa 3200003, Israel.; Otto von Guerricke University, Institute for Physics, Universit{\"a}tsplatz. 2, Magdeburg 39106, Germany.; SAWLab Saxony, Leibniz Institute for Solid State and Materials Research Dresden, Helmholtzstr. 20, Dresden 01069, Germany.},
	author = {Roudini, Mehrzad and Manuel Rossell{\'o}, Juan and Manor, Ofer and Ohl, Claus-Dieter and Winkler, Andreas},
	cois = {Declaration of Competing Interest The authors declare that they have no known competing financial interests or personal relationships that could have appeared to influence the work reported in this paper.},
	copyright = {Copyright {\copyright}2023 The Author(s). Published by Elsevier B.V. All rights reserved.},
	crdt = {2023/07/29 18:04},
	date = {2023 Aug},
	dep = {20230724},
	doi = {10.1016/j.ultsonch.2023.106530},
	edat = {2023/07/30 01:06},
	issn = {1873-2828 (Electronic); 1350-4177 (Print); 1350-4177 (Linking)},
	jid = {9433356},
	journal = {Ultrason Sonochem},
	jt = {Ultrasonics sonochemistry},
	language = {eng},
	lid = {S1350-4177(23)00242-0 {$[$}pii{$]$}; 10.1016/j.ultsonch.2023.106530 {$[$}doi{$]$}; 106530},
	lr = {20230810},
	mhda = {2023/07/30 01:07},
	month = {Aug},
	oto = {NOTNLM},
	own = {NLM},
	pages = {106530},
	phst = {2023/05/06 00:00 {$[$}received{$]$}; 2023/07/07 00:00 {$[$}revised{$]$}; 2023/07/19 00:00 {$[$}accepted{$]$}; 2023/07/30 01:07 {$[$}medline{$]$}; 2023/07/30 01:06 {$[$}pubmed{$]$}; 2023/07/29 18:04 {$[$}entrez{$]$}; 2023/07/24 00:00 {$[$}pmc-release{$]$}},
	pii = {S1350-4177(23)00242-0; 106530},
	pl = {Netherlands},
	pmc = {PMC10407539},
	pmcr = {2023/07/24},
	pmid = {37515911},
	pst = {ppublish},
	pt = {Journal Article},
	sb = {IM},
	status = {PubMed-not-MEDLINE},
	title = {Acoustic resonance effects and cavitation in SAW aerosol generation.},
	volume = {98},
	year = {2023}}

@article{Truby2016,
	author = {Truby, Ryan L. and Lewis, Jennifer A.},
	doi = {10.1038/nature21003},
	journal = {Nature},
	number = {7633},
	pages = {371--378},
	title = {Printing soft matter in three dimensions},
	volume = {540},
	year = {2016}}

@article{Hulme2024,
	author = {Hulme, James and Sakhaei, Amir Hosein and Shafiee, Mahmood},
	doi = {10.1016/j.matpr.2023.02.278},
	journal = {Materials Today: Proceedings},
	pages = {30--37},
	title = {Mechanical analysis and additive manufacturing of {3D}-printed lattice materials for bone scaffolds},
	volume = {101},
	year = {2024}}

@article{advanced2025electrospray,
	author = {Kang, Chaeyeong and Roh, Seokbeom and Park, Hyunjun and Lee, Sugeun and Lee, Taeha and Lee, Sang Won and Kim, Hansung and Park, Jinsung and Lee, Gyudo and Park, Insu},
	doi = {10.1002/advs.202510905},
	journal = {Advanced Science},
	number = {3},
	pages = {e10905},
	title = {Needle-free, liquid metal-embedded electrospray deposition system for controlled microdroplet printing},
	volume = {13},
	year = {2026}}

@article{deGans2004,
	author = {{de Gans}, Berend-Jan and Duineveld, Paul C. and Schubert, Ulrich S.},
	doi = {10.1002/adma.200300385},
	journal = {Advanced Materials},
	number = {3},
	pages = {203--213},
	title = {Inkjet printing of polymers: state of the art and future developments},
	volume = {16},
	year = {2004}}

@article{You2024,
	author = {You, Kejia and Sun, Yapeng and Li, Yan and Qian, Weipeng and Zhu, Jizhu and Cao, Xu},
	doi = {10.1002/smll.202402638},
	journal = {Small},
	number = {27},
	pages = {2402638},
	title = {On-demand picoliter-level-droplet inkjet printing for micro fabrication and functional applications},
	volume = {20},
	year = {2024}}

@article{Zhao2022,
	author = {Zhao, Siyu and Liu, Zhaomiao and Wang, Ju and Pang, Yan and Xue, Sen and Li, Mengqi},
	doi = {10.1063/5.0118015},
	journal = {Physics of Fluids},
	number = {11},
	pages = {112012},
	title = {Formation of high-viscosity micro-droplets in {T}-channels with neck structure induced by surface acoustic waves},
	volume = {34},
	year = {2022}}

@article{heinrich2019,
	author = {Heinrich, Marcel Alexander and Liu, Wanjun and Jimenez, Andrea and Yang, Jingzhou and Akpek, Ali and Liu, Xiao and Pi, Qingmeng and Mu, Xuan and Hu, Ning and Schiffelers, Raymond Michel and Prakash, Jai and Xie, Jingwei and Zhang, Yu Shrike},
	doi = {10.1002/smll.201805510},
	journal = {Small},
	number = {23},
	pages = {1805510},
	title = {{3D} bioprinting: from benches to translational applications},
	volume = {15},
	year = {2019}}

@article{Alvarez2008,
	author = {Alvarez, Marian and Friend, James R. and Yeo, Leslie Y.},
	doi = {10.1088/0957-4484/19/45/455103},
	journal = {Nanotechnology},
	number = {45},
	pages = {455103},
	title = {Rapid generation of protein aerosols and nanoparticles via surface acoustic wave atomization},
	volume = {19},
	year = {2008}}

@article{Turkoz2018,
	author = {T{\"u}rk{\"o}z, Emre and Perazzo, Antonio and Kim, Hyoungsoo and Stone, Howard A. and Arnold, Craig B.},
	doi = {10.1103/PhysRevLett.120.074501},
	journal = {Physical Review Letters},
	pages = {074501},
	title = {Impulsively induced jets from viscoelastic films for high-resolution printing},
	volume = {120},
	year = {2018}}

@article{Theodorakos2019,
	author = {Theodorakos, Ioannis and Kalaitzis, Alexandros and Makrygianni, Michaela and Tsakona, Dimitra and Kyriakidis, Nikolaos and Giannakopoulos, Stavros},
	doi = {10.1002/adem.201800605},
	journal = {Advanced Engineering Materials},
	number = {4},
	pages = {1800605},
	title = {Laser-induced forward transfer of high-viscous, non-{N}ewtonian silver nanoparticle inks: jet dynamics and temporal evolution of the printed droplet study},
	volume = {21},
	year = {2019}}

@article{serra2019,
	author = {Serra, Pere and Piqu{\'e}, Alberto},
	doi = {10.1002/admt.201800099},
	journal = {Advanced Materials Technologies},
	number = {1},
	pages = {1800099},
	title = {Laser-induced forward transfer: fundamentals and applications},
	volume = {4},
	year = {2019}}

@book{pique2003,
	address = {San Diego, CA},
	editor = {Piqu{\'e}, Alberto and Chrisey, Douglas B.},
	isbn = {978-0-12-174231-0},
	publisher = {Academic Press},
	title = {Direct-Write Technologies for Rapid Prototyping Applications: Sensors, Electronics, and Integrated Power Sources},
	year = {2002}}

@article{Foresti2018,
	author = {Foresti, Daniele and Kroll, Katharina T. and Amissah, Robert and Sillani, Francesco and Homan, Kimberly A. and Poulikakos, Dimos and Lewis, Jennifer A.},
	doi = {10.1126/sciadv.aat1659},
	journal = {Science Advances},
	number = {8},
	pages = {eaat1659},
	title = {Acoustophoretic printing},
	volume = {4},
	year = {2018}}

@article{KekeChen2021,
	author = {Chen, Keke and Jiang, Erhui and Wei, Xiaoyun and Xia, Yu and Wu, Zezheng and Gong, Zhiyi and Shang, Zhengjun and Guo, Shishang},
	doi = {10.1039/D1LC00003A},
	journal = {Lab on a Chip},
	number = {8},
	pages = {1604--1612},
	title = {The acoustic droplet printing of functional tumor microenvironments},
	volume = {21},
	year = {2021}}

@article{HongyiChen2025,
	author = {Chen, Hongyi and Hardwick, James and Gao, Lei and Martinez Plasencia, Diego and Subramanian, Sriram and Hirayama, Ryuji},
	doi = {10.1063/5.0271688},
	journal = {Applied Physics Reviews},
	number = {3},
	pages = {031305},
	title = {Acoustics in additive manufacturing: a path toward contactless, scalable, and high-precision manufacturing},
	volume = {12},
	year = {2025}}

@article{Li2020,
	author = {Li, Jie and Biroun, Mehdi H. and Tao, Ran and Wang, Yong and Torun, Hamdi and Xu, Na and Rahmati, Mohammad and Li, Yifan and Gibson, Desmond and Fu, Chen and Luo, Jingting and Dong, Linxi and Xie, Jin and Fu, Yongqing},
	doi = {10.1088/1361-6463/ab8f50},
	journal = {Journal of Physics D: Applied Physics},
	number = {35},
	pages = {355402},
	title = {Wide range of droplet jetting angles by thin-film based surface acoustic waves},
	volume = {53},
	year = {2020}}

@article{Friend2011,
	author = {Friend, James and Yeo, Leslie Y.},
	doi = {10.1103/RevModPhys.83.647},
	journal = {Reviews of Modern Physics},
	number = {2},
	pages = {647--704},
	title = {Microscale acoustofluidics: microfluidics driven via acoustics and ultrasonics},
	volume = {83},
	year = {2011}}

@article{YeoFriend2014,
	author = {Yeo, Leslie Y. and Friend, James R.},
	doi = {10.1146/annurev-fluid-010313-141418},
	journal = {Annual Review of Fluid Mechanics},
	pages = {379--406},
	title = {Surface acoustic wave microfluidics},
	volume = {46},
	year = {2014}}

@article{Tan2009,
	author = {Tan, Ming K. and Friend, James R. and Yeo, Leslie Y.},
	doi = {10.1103/PhysRevLett.103.024501},
	journal = {Physical Review Letters},
	number = {2},
	pages = {024501},
	title = {Interfacial jetting phenomena induced by focused surface vibrations},
	volume = {103},
	year = {2009}}

@article{Bhattacharjee2011,
	author = {Bhattacharjee, P. K. and McDonnell, A. G. and Prabhakar, R. and Yeo, L. Y. and Friend, J.},
	doi = {10.1088/1367-2630/13/2/023005},
	journal = {New Journal of Physics},
	pages = {023005},
	title = {Extensional flow of low-viscosity fluids in capillary bridges formed by pulsed surface acoustic wave jetting},
	volume = {13},
	year = {2011}}

@article{Castro2018,
	author = {Castro, Jasmine O. and Ramesan, Shwathy and Rezk, Amgad R. and Yeo, Leslie Y.},
	doi = {10.1039/C7SM02534C},
	journal = {Soft Matter},
	number = {27},
	pages = {5721--5727},
	title = {Continuous tuneable droplet ejection via pulsed surface acoustic wave jetting},
	volume = {14},
	year = {2018}}

@article{connacher2018,
	author = {Connacher, William and Zhang, Naiqing and Huang, An and Mei, Jiyang and Zhang, Shuai and Gopesh, Tilvawala and Friend, James},
	doi = {10.1039/C8LC00112J},
	journal = {Lab on a Chip},
	number = {14},
	pages = {1952--1996},
	title = {Micro/nano acoustofluidics: materials, phenomena, design, devices, and applications},
	volume = {18},
	year = {2018}}

@article{friend2023acoustofluidics,
	author = {Friend, James},
	doi = {10.3389/facou.2023.1261027},
	journal = {Frontiers in Acoustics},
	pages = {1261027},
	title = {Acoustofluidics},
	volume = {1},
	year = {2023}}

@article{Mei:aa,
	author = {Mei, Jiyang and Zhang, Naiqing and Friend, James},
	doi = {10.3791/61013},
	journal = {Journal of Visualized Experiments},
	pages = {e61013},
	title = {Fabrication of surface acoustic wave devices on lithium niobate},
	volume = {160},
	year = {2020}}

@article{Rufo2022,
	author = {Rufo, Joseph and Cai, Feiyan and Friend, James and Wiklund, Martin and Huang, Tony Jun},
	doi = {10.1038/s43586-022-00109-7},
	journal = {Nature Reviews Methods Primers},
	pages = {30},
	title = {Acoustofluidics for biomedical applications},
	volume = {2},
	year = {2022}}

@article{Zhou2014,
	author = {Zhou, J. and Pang, H. F. and Garcia-Gancedo, L. and Iborra, E. and Clement, M. and De Miguel-Ramos, M. and Jin, H. and Luo, J. K. and Smith, S. and Dong, S. R. and Wang, D. M. and Fu, Y. Q.},
	doi = {10.1007/s10404-014-1456-1},
	journal = {Microfluidics and Nanofluidics},
	number = {4},
	pages = {537--548},
	title = {Discrete microfluidics based on aluminum nitride surface acoustic wave devices},
	volume = {18},
	year = {2015}}

@article{LeiHu2020Review,
	author = {Lei, Yulin and Hu, Hong},
	doi = {10.1063/5.0014768},
	journal = {Biomicrofluidics},
	number = {6},
	pages = {061505},
	title = {{SAW}-driven droplet jetting technology in microfluidic: a review},
	volume = {14},
	year = {2020}}

@article{Lei2020Actuators,
	author = {Lei, Yulin and Hu, Hong and Chen, Jian and Zhang, Peng},
	doi = {10.3390/act9010005},
	journal = {Actuators},
	number = {1},
	pages = {5},
	title = {Microfluidic jetting deformation and pinching-off mechanism in capillary tubes by using traveling surface acoustic waves},
	volume = {9},
	year = {2020}}

@article{Biroun2020,
	author = {Biroun, Mehdi H. and Rahmati, Mohammad and Jangi, Mehdi and Chen, Baixin and Fu, Yong Qing},
	doi = {10.1016/j.ijmultiphaseflow.2020.103545},
	journal = {International Journal of Multiphase Flow},
	pages = {103545},
	title = {Numerical and experimental investigations of interdigital transducer configurations for efficient droplet streaming and jetting induced by surface acoustic waves},
	volume = {136},
	year = {2021}}

@article{connacher2020,
	author = {Connacher, William and Orosco, Jeremy and Friend, James},
	doi = {10.1103/PhysRevLett.125.184504},
	journal = {Physical Review Letters},
	pages = {184504},
	title = {Droplet ejection at controlled angles via acoustofluidic jetting},
	volume = {125},
	year = {2020}}

@article{Ning2023,
	author = {Ning, Jia and Lei, Yulin and Hu, Hong and Gai, Chenhui},
	doi = {10.3390/mi14081543},
	journal = {Micromachines},
	number = {8},
	pages = {1543},
	title = {A comprehensive review of surface acoustic wave-enabled acoustic droplet ejection technology and its applications},
	volume = {14},
	year = {2023}}

@article{Orosco2022,
	author = {Orosco, Jeremy and Friend, James},
	doi = {10.1103/PhysRevE.106.045101},
	journal = {Physical Review E},
	number = {4},
	pages = {045101},
	title = {Modeling fast acoustic streaming: steady-state and transient flow solutions},
	volume = {106},
	year = {2022}}

@article{Derby2010,
	author = {Derby, Brian},
	doi = {10.1146/annurev-matsci-070909-104502},
	journal = {Annual Review of Materials Research},
	pages = {395--414},
	title = {Inkjet printing of functional and structural materials: fluid property requirements, feature stability, and resolution},
	volume = {40},
	year = {2010}}

@article{Khalil2024,
	author = {Khalil, Shaheer Mohiuddin and Ali, Shahzaib and Nguyen, Vu Dat and Cho, Dae-Hyun and Byun, Doyoung},
	doi = {10.1002/aisy.202400621},
	journal = {Advanced Intelligent Systems},
	number = {1},
	pages = {2400621},
	title = {High-precision drop-on-demand printing of charged droplets on nonplanar surfaces with machine learning},
	volume = {7},
	year = {2025}}

@article{Eggers2008,
	author = {Eggers, Jens and Villermaux, Emmanuel},
	doi = {10.1088/0034-4885/71/3/036601},
	journal = {Reports on Progress in Physics},
	number = {3},
	pages = {036601},
	title = {Physics of liquid jets},
	volume = {71},
	year = {2008}}

@article{Yarin2006,
	author = {Yarin, Alexander L.},
	doi = {10.1146/annurev.fluid.38.050304.092144},
	journal = {Annual Review of Fluid Mechanics},
	pages = {159--192},
	title = {Drop impact dynamics: splashing, spreading, receding, bouncing{\ldots}},
	volume = {38},
	year = {2006}}

@article{Josserand2016,
	author = {Josserand, Christophe and Thoroddsen, Sigurdur T.},
	doi = {10.1146/annurev-fluid-122414-034401},
	journal = {Annual Review of Fluid Mechanics},
	pages = {365--391},
	title = {Drop impact on a solid surface},
	volume = {48},
	year = {2016}}

@article{Huang2021,
	author = {Huang, An and Connacher, William and Stambaugh, Mark and Zhang, Naiqing and Zhang, Shuai and Mei, Jiyang and Jain, Aditi and Alluri, Sravya and Leung, Vincent and Rajapaksa, Anushi E. and Friend, James},
	doi = {10.1039/D0LC01008A},
	journal = {Lab on a Chip},
	number = {7},
	pages = {1352--1363},
	title = {Practical microcircuits for handheld acoustofluidics},
	volume = {21},
	year = {2021}}

@article{Soltman2008,
	author = {Soltman, Daniel and Subramanian, Vivek},
	doi = {10.1021/la7026847},
	journal = {Langmuir},
	number = {5},
	pages = {2224--2231},
	title = {Inkjet-printed line morphologies and temperature control of the coffee ring effect},
	volume = {24},
	year = {2008}}

@article{Stringer2010,
	author = {Stringer, Jonathan and Derby, Brian},
	doi = {10.1021/la101296e},
	journal = {Langmuir},
	number = {12},
	pages = {10365--10372},
	title = {Formation and stability of lines produced by inkjet printing},
	volume = {26},
	year = {2010}}

@article{Lohse2022,
	author = {Lohse, Detlef},
	doi = {10.1146/annurev-fluid-022321-114001},
	journal = {Annual Review of Fluid Mechanics},
	pages = {349--382},
	title = {Fundamental fluid dynamics challenges in inkjet printing},
	volume = {54},
	year = {2022}}

@article{Jentsch2021,
	author = {Jentsch, Stefan and Nasehi, Ramin and Kuckelkorn, Christoph and Gundert, Benedikt and Aveic, Sanja and Fischer, Horst},
	doi = {10.1002/smtd.202000971},
	journal = {Small Methods},
	number = {6},
	pages = {e2000971},
	title = {Multiscale {3D} bioprinting by nozzle-free acoustic droplet ejection},
	volume = {5},
	year = {2021}}

\end{document}